# Charge transport in monolayers of metal nanoparticles

Lianhua Zhang,[a] Ji'an Chen,[b] Fei Liu,[b] Zhengyang Du,[b] Yilun Jiang,[b] Min Han[b]*

Email：sjhanmin@nju.edu.cn

[a]*National Laboratory of Solid State Microstructures, School of Physics, and Collaborative Innovation Centre of Advanced Microstructures, Nanjing University, 210093 Nanjing, China.*

[b]*National Laboratory of Solid State Microstructures, College of Engineering and Applied Sciences, and Collaborative Innovation Centre of Advanced Microstructures, Nanjing University, 210093 Nanjing, China.*

**Abstract**

Two-dimensional (2D) nanoparticle films are a new class of materials with interesting physical properties and applications ranging from nanoelectronics to sensing and photonics. The importance of conducting nanoparticle films makes the fundamental understanding of their charge transport extremely important for materials and process design. Various hopping and transport mechanisms have been proposed and the nanoparticle monolayer is consistent with the electrical equivalent RC circuit, but their theoretical methods are limited to the model of the single electron tunneling between capacitively coupled nanoparticles with a characteristic time constant $\tau=RC$ and the conductivity $\sigma(T)$ of thin film $\sigma(T) \propto \exp(-E_T/k_B T)$ is the experimental conductivity, which cannot be deduced from these theoretical models. It is also unclear that how the specific process of electron transpot is affected by temperature. So, nowadays the electron dynamics of thin film cannot be understood fundamentally. Here, we develop an analytical theory based on the model of Sommerfeld, backed up by Monte-Carlo simulations, that predicts the process of charge transport and the effect of temperature on the electron transport in the thin film. In this paper two different nanoparticle models were built to cope with different types of morphology: triangular array and rectangular array. The transport properties of these different kinds of arrays including 2D ordered nanoparticle arrays with/without local structural disorder and 2D gradient nanoparticle arrays were investigated at different temperatures. For 2D well-ordered nanoparticle array without local structural disorder, the *I-V* curves are non-linear and highly symmetric. However, the *I-V* curves are also non-linear but highly asymmetric in 2D gradient nanoparticle array, which is very similar to the characteristics of diode. For these properties, 2D gradient nanoparticle array could have a variety of applications in the future. Furthermore, we find that these nanoparticle arrays ranging from void-filled networks to well-ordered superlattices show clear voltage thresholds due to Coulomb blockade, and temperature-dependent conduction indicative of quantum tunneling.

**Keywords**: metal nanoparticle; electron transport; quantum tunneling; gradient array; local structural disorder; close-packed monolayer

## 1. Introduction

Nanostructured materials have raised substantial interest in research and industry [1-4]. They are used or discussed for a wide range of applications [5], like in optical devices [6], or in electronic [7] and magnetic architectures [8, 9]. Furthermore, nanoparticles and their assemblies are interesting model systems for investigations of the physical and chemical properties on the nanometer scale [10, 11]. The electronic transport properties of these systems are the subject of numerous publications [4, 6, 7, 12-15]. A vast number of experimental and theoretical results dealing with the electron transport properties of nanoparticle arrays have been published [9, 13, 14, 16-24]. The transport characteristics of nanoparticle films are strongly affected by the distribution of nanoparticle on the substrate [25]. Theoretical approaches dealing with tunneling transport so far have concentrated on local charge disorder alone [22]. The combined types of disorder are not considered, even though large differences between spatially ordered and disorder structures might be expected due to the exponential dependence of local tunneling resistances on the interparticle spacings. A wide variety of disordered systems, such as the disordered superconductor [26-29], charge-density waves pinned by impurities [30], Wigner crystals [31,32] in semiconductors with charge impurities and colloids flowing over rough surfaces [33], exhibits threshold behavior and nonlinear response to an applied voltage [34].

The gradient nanoparticle assembly [35–46], having different properties from 2D well-distributed nanoparticle array, can be used as a combinatorial tool for manipulation and investigation of physicochemical properties of materials systematically at the micro/nanoscale [42-46]. For example, the gradient nanoparticle array can provide various advantages for exploring surface enhanced Raman scattering substrates with optimal conformations for highly sensitive and reproducible detection of specific molecules [37]. In addition, the gradient nanoparticle array with tunable scales and gradient steepness can be envisaged to have new phenomena and properties associated with the gradually changing interfaces and multiscale hybridization, and open new opportunities for designing optoelectronic nanodevices and nanosystems. So, the understanding of electron dynamics in the gradient nanoparticle film is important. Nowadays a systematic study about the electronic transport mechanism of 2D gradient nanoparticle film has not been done.

Many simple theoretical models have to be developed [13-20] to investigate thin film with the single electron tunneling between capacitively coupled nanoparticles with a characteristic time constant $\tau=RC$. Various hopping and transport mechanisms have been proposed: the simple thermally activated hopping for ordered systems [21] and variable range hopping (VRH) for granular metals with highly disordered systems [22]. For thermally activated transport, the conductivity $\sigma$ is often described with an Arrhenius equation of the form $\sigma(T) \propto \exp(-E_T/k_BT)$, where $\sigma(T)$ is the experimental conductivity. Here $E_T$ is the activation energy and $k_B$ is the Boltzmann constant. For the VRH-driven transport, a different correlation was proposed: $\sigma(T) \propto \exp(-T_0/T)^{1/2}$. Here $T_0$ can be interpreted as the activation temperature. While the first mechanism describes the charge transfer between spatially adjacent particles, the VRH mechanism considers the transfer between the energetically closest particles states.

In this paper, we develop an analytical theory based on the model of electron of Sommerfeld, backed up the Monte-Carlo simulations, that predict the process of charge transport in conducting nanoparticle films. The effect of temperature on the characteristics of current-voltage was also considered in my model. The dynamic behavior along the conduction paths and their topology can be analyzed quantitatively. In our theory the single-electron tunneling dynamics and the Coulomb blockade effect are used to investigate the electronic transport mechanism.

## 2. The new model of charge transport in the nanoparticle array

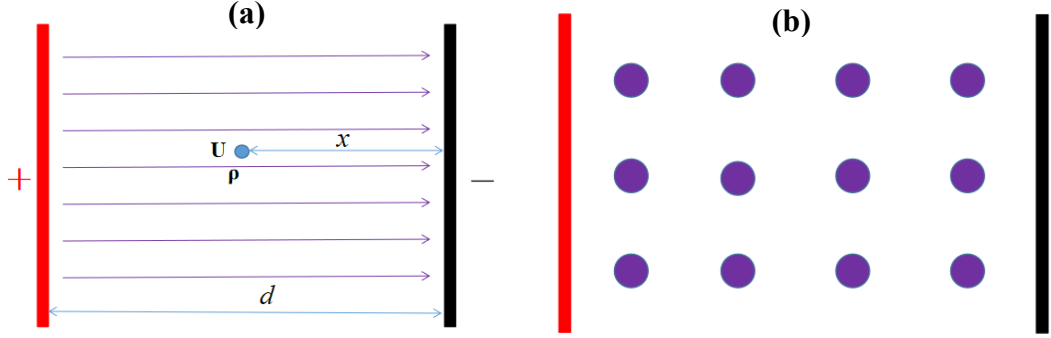

FIG. 1.(a) Parallel plate electrode system; (b) The schematic of nanoparticle array

Fig. 1a shows the parallel plate electrode system. In this system the relationship between the electric potential and the space charge density is following the Poisson equation:

$$\frac{d^2U}{dx^2} = -\frac{\rho}{\varepsilon_0} \tag{1}$$

$\rho$ is the space charge density; $U$ is the electric potential.

In some point of the parallel plate electrode system, the relationship between the velocity of electron and the electric potential can be described:

$$\vec{\upsilon} = \sqrt{\frac{2eU}{m}}\vec{I} \tag{2}$$

$\vec{I}$ is the unit vector. So the current density is

$$\vec{j} = -\rho\vec{\upsilon} = -\rho\sqrt{\frac{2eU}{m}}\vec{I} \tag{3}$$

The space charge density is

$$\rho = -\vec{j}\sqrt{\frac{m}{2eU}}\vec{I} \tag{4}$$

So the Poisson equation can be described:

$$\frac{d^2U}{dx^2} = -\frac{\vec{j}}{\varepsilon_0}\sqrt{\frac{m}{2eU}}\vec{I} \tag{5}$$

In this system the electric potential of the cathode is following $U=0$ and $\frac{dU}{dx}=0$, the current density can be expressed after carrying out two integrations.

$$\vec{j} = \frac{4\varepsilon_0}{9}\sqrt{\frac{2e}{m}}\frac{U^{3/2}}{x^2}\vec{I} \tag{6}$$

At some point of the parallel plate electrode system, $x=d_a$, $U=U_a$

$$\vec{j}_a = \frac{4\varepsilon_0}{9}\sqrt{\frac{2e}{m}}\frac{U_a^{3/2}}{d_a^2}\vec{I} \tag{7}$$

When dealing with isolated metal nanoparticles separated as islands by poorly permeable tunneling barriers distributed between two outer electrodes (see Fig. 2b), the number of electrons in these particles is always integer and will change if a certain voltage is applied. Since only electrons can enter or leave the nanoparticle as an entity, their number is also integer. Thus, the charge transport in arrays of conducting nanoparticles can be controlled by the external voltage (see Fig. 1b). If a nanoparticle is charged by only one excess electron, its electronic potential will rise to prevent further charging. This process is characterized as "single electron tunneling" — Coulomb blockade phenomena in nanostructures (see Fig. 3).

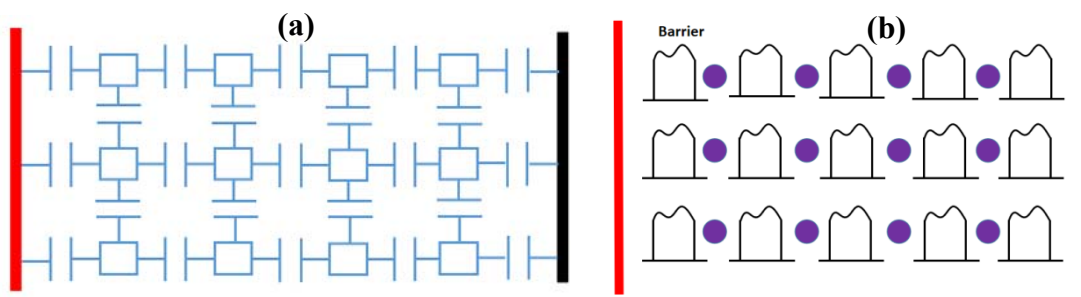

FIG. 2 (a) The traditional RC model of nanoparticle array; (b) The new model of nanoparticle array

For the nanoparticle array, the traditional model of nanoparticle array that is proposed by Middleton and Wingreen (MW) [22] can be described as following: the nanoparticles are treated as capacitively coupled conductions with charges allowed to tunnel between neighboring nanoparticles (see Fig. 2a). In our model each nanoparticle is independent, so it can be treated as a discrete phase model that an electron in some nanoparticle passing through a barrier. Base on the model of Sommerfeld, electrons in the interior of metallic nanoparticle are moving in the constant potential field. When the electron of one metallic nanoparticle reach the surface of this nanoparticle, an attractive force is generated to prevent the electron escaping from this metallic nanoparticle. This force can only work in a very short distance of nanoparticle surface and the barrier is produced by this force. If an appropriate electric field is applied, the electron will have positive energy on the surface of nanoparticle, then the electron has a certain chance to leave from this nanoparticle. So the barrier is dominated by the attraction force of process when electron transmits from one nanoparticle to an adjacent nanoparticle. So the transmission probability without an applied voltage is

$$P \sim e^{-2\sqrt{2mE_g}L/\hbar} \tag{8}$$

$E_g$ is energy gap, $L$ is the barrier width, $m$ is the effective mass of the electron and

$$E_g = V - E_a \tag{9}$$

In equation (9) $V$ is the barrier height, which is related to the size of nanoparticle, and $E_a$ is the kinetic energy of the electron. This means the charge transport on the microscopic scale is essentially affected by structural variation.

In the nanoparticle array each nanoparticle is affected by single-electron tunneling dynamics and the Coulomb blockade effect. So when the electric potential of nanoparticle is $U_a$, the probability density of single electron transporting from this nanoparticle with an extra electron to the neighboring nanoparticle is

$$P \sim \frac{U_a^{3/2} \vec{I}}{d_a^2} e^{-2\sqrt{2mE_g}L/\hbar} \tag{10}$$

When the applied voltage between two electrodes is $U$, the uniform electric filed $E$ is present

$$E = U/d \tag{11}$$

where $d$ is the distance between two electrodes. The electric potential $U_a$ of some nanoparticle is

$$U_a = U d_a / d \tag{12}$$

$d_a$ is the distance between some nanoparticle and cathode. So the transmission probability with applied voltage is

$$P \sim \frac{(U/d)^{3/2} \vec{I}}{d_a^{1/2}} e^{-2\sqrt{2mE_g}L/\hbar} \tag{13}$$

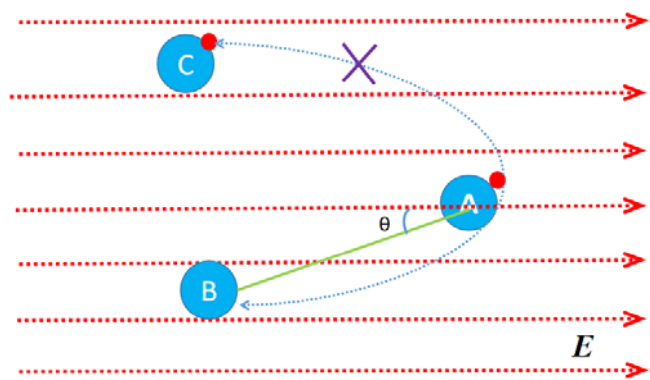

Fig. 3 The schematic of single electron tunneling and the Coulomb blockade effect

Under the condition of single-electron tunneling and the Coulomb blockade effect, the transmission probability is

$$P \sim \frac{(U/d)^{3/2} \cos\theta}{d_a^{1/2}} e^{-2\sqrt{2mE_g}L/\hbar} \tag{14}$$

$\theta$ is the relative angle between electric field line and the connecting line of two adjacent nanoparticles, where electron could transmit from one nanoparticle to the neighboring nanoparticle.

## 3. Modeling nanoparticle assemblies

Metal nanoparticle films can be produced by various methods [2]. Their morphologies include regular closed-packed arrays [27, 28] and nano-networks with well-organized cells [14, 29], as well as multicellular fractal structures [30] and glassy disordered deposits [31]. Each type of assembly exhibits different properties. We will precisely delineate the roles played by the different types of ordered and disordered 2D nanoparticle arrays in the numerical modeling of charge transport in metal nanoparticle films, which are conducting via single-electron tunneling mechanisms.

Since much larger contiguous paths of nanoparticles are found in the array with long-range order than in the void-ridden arrays, the charge transport behavior is directly relevant to the probability $P(n)$ of a nanoparticle being in a linear segment of $n$ particles uninterrupted by voids. From equation (14), we can obtain the behavior of charge transport implicates direct, interparticle quantum tunneling as the conduction mechanism and the series-parallel combination of many tunneling paths from particle to particle throughout the array. So the model of rectangular array is built to demonstrate the structure-dynamics interdependence in the nanoparticle film using theoretical arguments and quantitative analysis (see Fig. 4, Fig.5 and Fig. 6). If the 2D nanopartcle array has large void fraction, neighboring voids produce bottlenecks, locally cutting off the transverse correlation length. In the extreme case, conduction is reduced to several parallel 1D channels. In this situation the model of triangular array is proposed (see Fig. 8). In these two models the arrays consist of small normal metal nanoparticles linked by tunnel barriers, in which transport occurs through the stochastic flow of discrete charges.

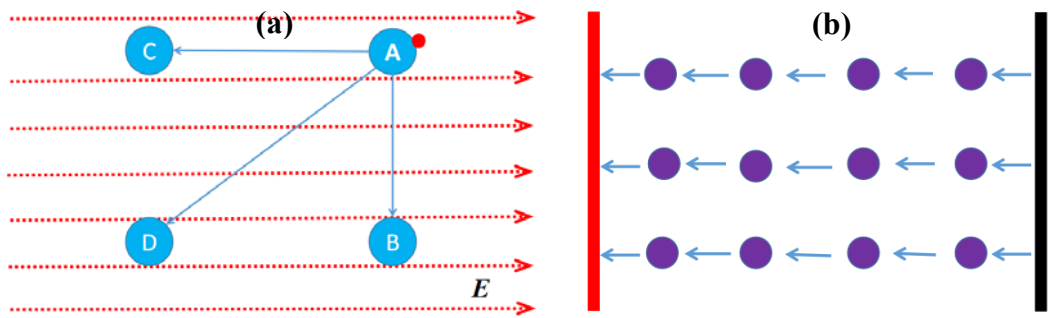

Fig.4 (a) The schematic of three representative processes of single-electron tunneling. (b)The schematic of electron transport path in the ordered nanoparticle array.

Regular close-packed array corresponding to the density balanced mono-layer metal nanoparticle film with few voids can be modeled on the ordered rectangular array of metal nanoparticle, which is shown in Fig. 1b. In the model we assume three representative processes of single-electron tunneling may occur in a uniform array with a constant inter-particle distance between neighboring particles, as shown in Fig. 4a. When particle B, C and D have no charges, particle A with an extra electron could transmit the electron to one of the three particles. The tunneling rates of the three processes can be given from equation (14).

$$P_{A \to B} \sim \frac{(U/d)^{3/2} \cos\theta_{AB}}{d_A^{1/2}} e^{-2\sqrt{2mE_g} L_{AB}/\hbar} \tag{15}$$

$$P_{A \to C} \sim \frac{(U/d)^{3/2} \cos\theta_{AC}}{d_A^{1/2}} e^{-2\sqrt{2mE_g} L_{AC}/\hbar} \tag{16}$$

$$P_{A \to D} \sim \frac{(U/d)^{3/2} \cos\theta_{AD}}{d_A^{1/2}} e^{-2\sqrt{2mE_g} L_{AD}/\hbar} \tag{17}$$

In a uniform array of nanoparticles, we can get the follow relationship

$$\theta_{AC} < \theta_{AD} < \theta_{AB} \cong 90° \tag{18}$$
$$L_{AB} = L_{AC} < L_{AD} \tag{19}$$

So the relationship among the tunneling rates of these processes can be expressed as

$$P_{A \to C} \gg P_{A \to D} \gg P_{A \to B} \tag{20}$$

Equation (20) implies that in a uniform array the charge moves only in 1D channels without any jumping of charge between adjacent channels. It means the most possible process from nanoparticle A to nanoparticle C may occur in the film (see Fig. 4a).

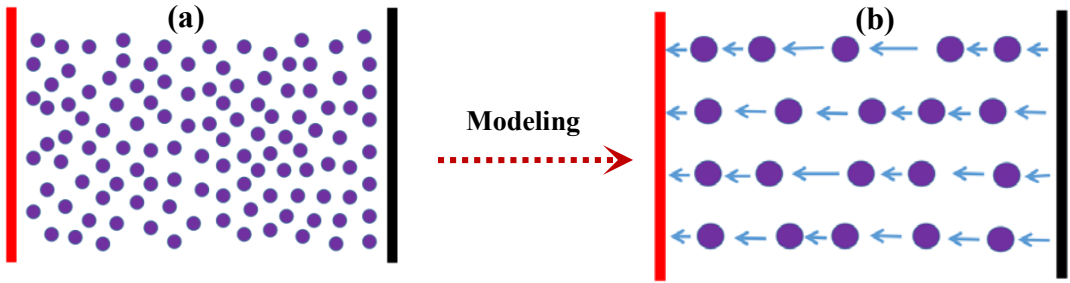

FIG. 5 (a) Well-packed nanoparticle monolayer with numerous voids. (b) The electron transport path in the disordered rectangular array with applied voltage.

The uniform array can be transformed into 1D particle chains with *n* particles in the simulation (see Fig. 4b). However, in order to reduce statistical fluctuations of charge flow, multichannel same chains of nanoparticles are used in this simulation.

For the disordered array (As shown in Fig. 5a), because of the different distance between every two adjacent nanoparticles in one channel, the charges in one channel do not synchronize with the flow of charge in adjacent channels. The charges are independently moving in the longitudinal direction. Since the channels have different amounts of flowing charge, multiple sets of nanoparticles that are different from each other are arranged to reduce the statistical fluctuations of the amount of electron in the simulation. So, the total current is a collective effect, which is the amount of electron in all channels (see Fig. 5b). This method can also be used in the rectangular gradient array (see Fig. 6b). The total current can be expressed as

$$I = I_1 + I_2 + I_3 + I_4 + I_5 + ...$$

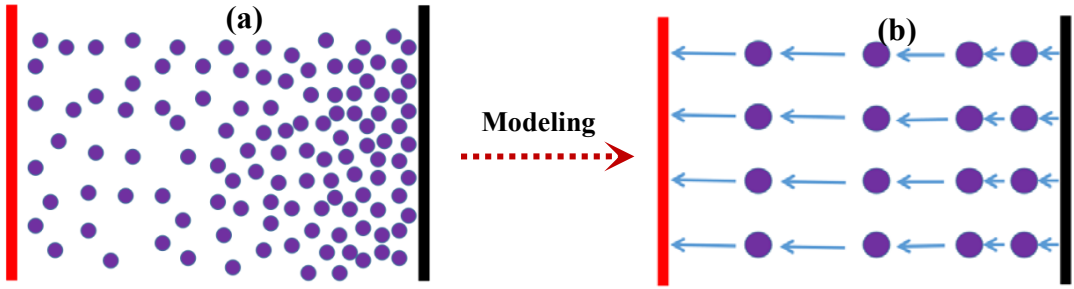

FIG. 6. (a) The rectangular gradient array represents the gradient nanoparticle film with numerous large voids. (b) The electron transport path in the rectangular gradient array with applied voltage.

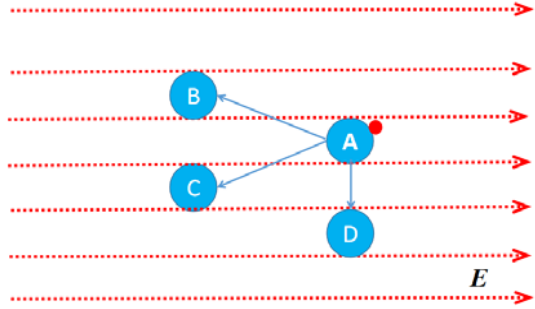

FIG. 7. The schematic of three representative processes of single-electron tunneling in the triangular array.

There are also three representative processes of single-electron tunneling in the model of triangular array, which is shown in Fig. 7.

$$P_{A \to B} \sim \frac{(U/d)^{3/2} \cos\theta_{AB}}{d_A^{1/2}} e^{-2\sqrt{2mE_g}L_{AB}/\hbar} \tag{21}$$

$$P_{A \to C} \sim \frac{(U/d)^{3/2} \cos\theta_{AC}}{d_A^{1/2}} e^{-2\sqrt{2mE_g}L_{AC}/\hbar} \tag{22}$$

$$P_{A \to D} \sim \frac{(U/d)^{3/2} \cos\theta_{AD}}{d_A^{1/2}} e^{-2\sqrt{2mE_g}L_{AD}/\hbar} \tag{23}$$

For triangular array,

$$\theta_{AB} = \theta_{AC} < \theta_{AD} \cong 90° \quad (24)$$

$$L_{AB} = L_{AC} = L_{AD} \quad (25)$$

From equation (21) to equation (25), we can obtain:

$$P_{A \to B} = P_{A \to C} \gg P_{A \to D} \quad (26)$$

Therefore, electron transmission could happen in two representative paths, the process from nanoparticle A to nanoparticle B and another process from nanoparticle A to nanoparticle C are shown in Fig. 7. There is a crossover from one path to the other which open, close, and shift position over time (see Fig. 8).

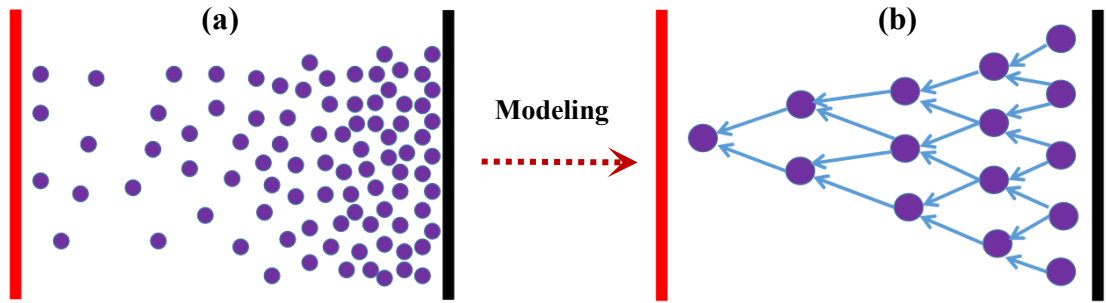

FIG. 8. (a) The triangular gradient array represents inhomogeneous film with void-filled networks. (b) The electron transport path in the triangular gradient array with applied voltage.

For the gradient array, the ring lattice with the uniform number density of metal nanoparticle, which is equal to the gradient array, can be used to investigate the characteristics of voltage-current. From the transmission probability, the number of electron transport path is decreasing when electron transport from cathode to anode (see Fig. 9).

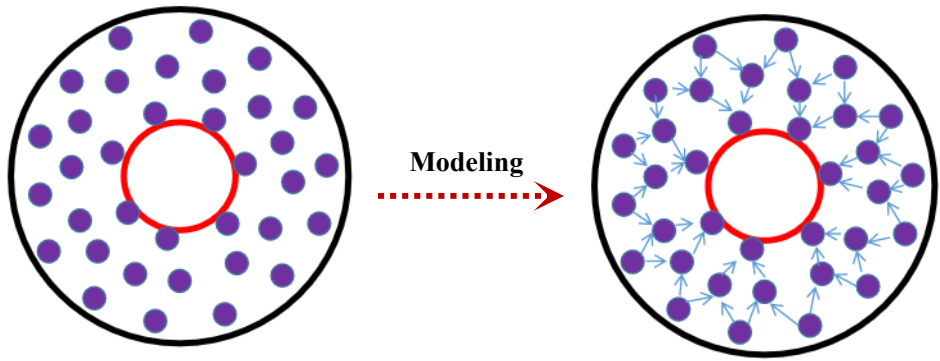

FIG. 9. (a) The ring array represents gradient array. (b) The electron transport path in the ring lattice with an applied voltage.

4. **Results and discussion**

The charge transport properties of 2D nanoparticle array are determined by many interrelated parameters, such as local structural disorder and gradient. An improper choice of these parameter may lead to no current, whereas their careful optimization can provide the nanoparticle array with the versatility to different characteristics of the current, symmetry and asymmetry, depending on the distribution of nanoparticle in the film.

Experimentally, various films with different types of order or disorder were fabricated through different methods. The distribution of nanoparticle is found to have the most critical effect on the transport characteristics of film. Considering the diversity on the experimentally fabricated nanoparticle arrays, two models are used in the simulation: triangular array and rectangular array, as illustrated in Fig. 1b and Fig. 8.

Firstly, we consider a rectangular ordered array of $N \times M$ nanoparticles. We introduce local structural disorder by assigning inconstant distance between the neighboring nanoparticles, as shown in Fig. 5. The coordinate of each nanoparticle can be described as following:

$$X(m) = (m^s - 0.5)L_x \tag{27}$$
$$Y(n) = (n - 0.5)L_y \tag{28}$$

where $s$ is the factor that would dominate the arrange of nanoparticle in the array. $L_x$ and $L_y$ are the unit length and the unit width respectively. When $s = 1.0$, the distance between the neighboring nanoparticles is always unchanged. But for $s = 1.1$ or $s = 1.2$, the distance of the neighboring nanoparticles is inconstant in the array, as shown in Fig. 6b, which is called a "rectangular gradient array". In the experiment, numerous voids may exist in the gradient array, which could give rise to the fluctuation of number density of nanoparticles. Therefore, we considered rectangular gradient array. In this type of array, the Y coordinates of nanoparticles remain unchanged. The oscillation amplitude of every nanoparticle in the array can be described as following:

$$A_0 = C_V L_x \tag{29}$$
$$A(n) = A_0 \sin(2\pi R) \tag{30}$$

So, the X-coordinates of nanoparticles in the array with local structural disorder are expressed as:

$$X(n) = (m^s - 0.5)L_x + C_V L_x \sin(2\pi R) \tag{31}$$

$C_V$ is a variable parameter $(0 \leq C_V < 1)$, $R$ is the random number $(0 \leq R \leq 1)$.

In the simulation the probability of electron transporting from one nanoparticle to an adjacent nanoparticle is

$$P \sim \frac{(U/d)^{3/2} \cos\theta}{d_a^{1/2}} e^{-2\sqrt{2mE_g}L/\hbar} \tag{32}$$

IF $P \geq R$, $R$ is the random number $(0 \leq R \leq 1)$, we consider this electron transport process can happen in the nanoparticle array.

The temperature is an important factor for the electron transport, so the influence of temperature should be taken into account in the simulation. With temperature increasing, the electrons in the nanoparticle could have enough thermal kinetic energies to pass through the barrier.

We suppose the kinetic energy of electron is $E_a$ when the temperature is $T$. When the temperature goes up to $T + \Delta T$, the energy of electron is

$$E'_a = E_a + k_B \Delta T \tag{33}$$

$k_B$ is the Boltzmann constant and the energy gap $E'_g$ is

$$E'_g = V - E' = E_g - k_B \Delta T \tag{34}$$

Then the transmission probability $P$ can be expressed as:

$$P \sim \frac{(U/d)^{3/2} \cos\theta}{d_a^{1/2}} e^{-2\sqrt{2m(E_g - k_B \Delta T)}L/\hbar} \tag{35}$$

## 4.1 The characteristics of current-voltage in the 2D ordered array

*4.1.1 The effect of temperature on the characteristics of current-voltage in the 2D ordered array*

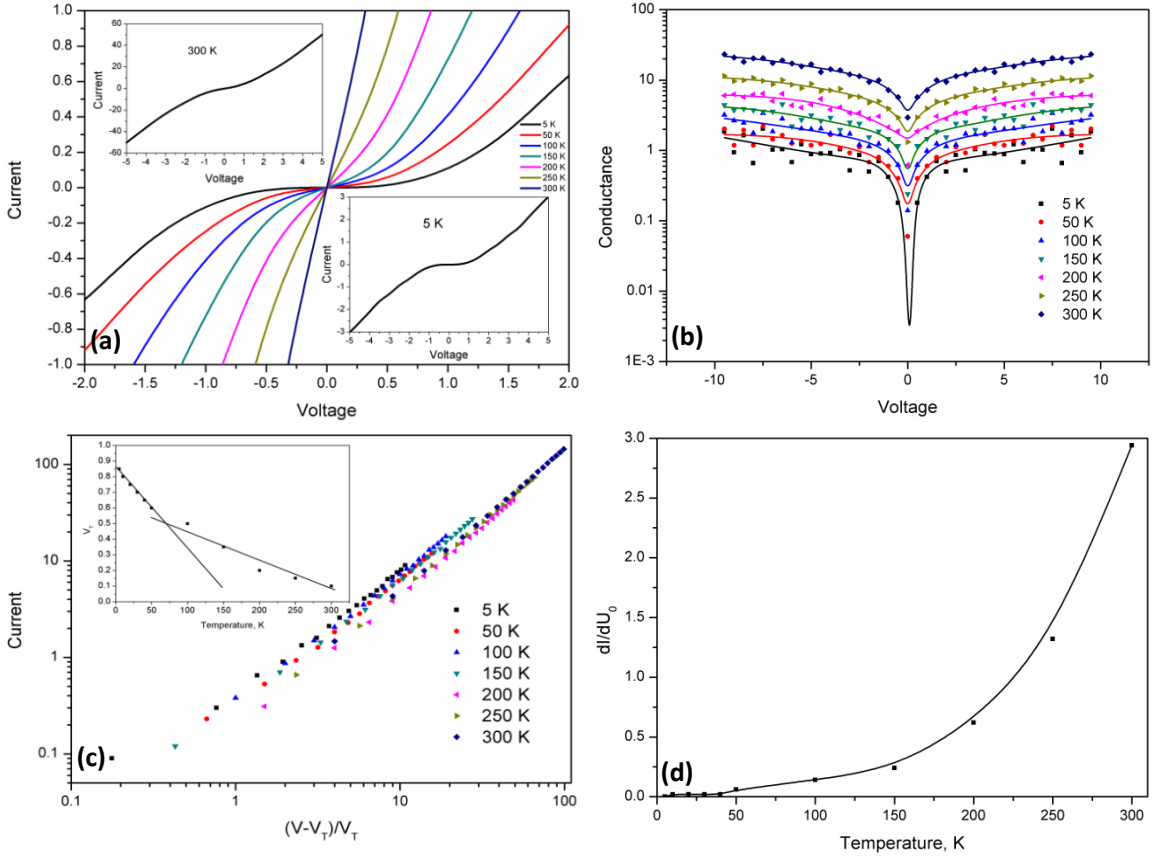

FIG. 10. (a) The *I-V* characteristics of the uniform rectangular array at seven temperatures for the same array. The upper inset shows the *I-V* curves of the same array taken with a larger bias range at 300 K, displaying almost ohmic conduction. The lower inset shows the *I-V* curves of the same array taken with a larger bias range at 5 K, displaying a nonlinear response. (b) The evolution of logarithmic plot of the differential conductance with temperature for the same sample. (c) The raw data is replotted as $[(V-V_T)/V_T]$ versus current. The inset shows the temperature dependence of the threshold voltage $V_T$ for the same array. (d) The development of zero bias conductance of the same array with temperature.

In this simulation we consider a rectangular ordered array of 10×10 nanoparticles and the unit length is $L_x = L_y = 1.0$. In order to probe the transport properties of the nanoparticle array, the *I-V* characteristics of the 2D array were calculated using numerical simulations at different temperatures. As shown in Fig. 10a, all of *I-V* curves are highly symmetric. A clear current blockade is present at 5K. While at room temperature the *I-V* curve is almost ohmic conduction. For low temperatures, no transport takes place up to a certain voltage that delivers the energy needed to overcome the Coulomb blockade. This so-called threshold voltage decreases with increasing temperature, which is very well visible in the logarithmic plot of the respective differential conductances in Fig. 10b. Even at room temperature, the characteristic is still slightly nonlinear. In consistent with other studies on 2D nanoparticle arrays, the current reduces as the temperature decreases. The nonlinear *I-V* characteristics exhibit a conspicuous threshold voltage at subambient temperatures. The *I-V* behavior of the nanoparticle array is analogous to a second order phase transition given by $I \sim [(V-V_T)/V_T]^\zeta$. There is in excellent agreement with the scaling law for the full temperature range of study (see Fig. 10c). Our data is also qualitatively in good agreement with the simulations of MW. We also observe scaling behavior above threshold voltage, which is predicted by the theory of MW and verified by experimental studies [16, 34].

The *I-V* curves in Fig. 10a exhibit a clear threshold voltage $V_T$, which is a characteristic of Coulomb blockade. At low temperatures, the linear decrease in $V_T$ with increasing temperature is similar to other 2D arrays studied [47, 48]. However, a sharp change in slope occurs around 50 K (see the upper inset of Fig. 10c). Above 50 K, the $V_T$ becomes a weak function of $T$. It is consistent with other 2D array studies [35]. Examining the differential conductance (see Fig. 10b and Fig. 10d) gives a deeper insight into the charge transport mechanism. For $V < V_T$, no current flows in the array and the values of conductance are zero near zero bias voltage at different temperatures. For $V > V_T$, the values of conductance increase rapidly at different temperatures and the enhancement of the current is attributed to the opening of many conducting paths through the sample, which contribute to the cooperative dynamical behavior of the charge transport. In the nanoparticle array, there are two types of electron emission: cold electron emission and thermal electron emission. Cold electron emission in some nanoparticle is depended on the electric potential and thermal electron emission is related to the temperature. When the applied voltage is low, electron can not escape from this nanoparticle. Moreover, when the temperature is limited within a certain range, there is no electron escaping from this nanoparticle in a low applied voltage ($V < V_T$). So the value of threshold voltage is related to a collective effect, which is related to cold electron emission and thermal electron emission.

In the experiment, the measured threshold voltage $V_{ET}$ are depended on the sensitivity of instrument, the size of the array, the spatial distribution of nanoparticles, the composition of nanoparticle, the number density of nanoparticles, the size of nanoparticle and so on. According to the quantum tunneling effect, a few electrons can transmit from cathode to anode when the applied voltage is very low. But for the voltage below a certain level, the current is too small to be distinguished from the conduction noise. So in the experiment the current is zero below some voltage. When the applied voltage is increasing, the potential of each nanoparticle in the array is enlarging. From equation (14), we can deduce that the quantum tunneling rate of electron from one nanoparticle to the neighboring nanoparticle is increasing rapidly. So numerous electrons can travel through the nanoparticle array. When the value of current reach the minimum value that the instrument can detect, the applied voltage is so-called experimental threshold voltage $V_{ET}$, which should be larger than the threshold voltage $V_T$ that the theory predicted.

*4.1.2 The effect of the unit length on the characteristics of current-voltage in the 2D ordered array*

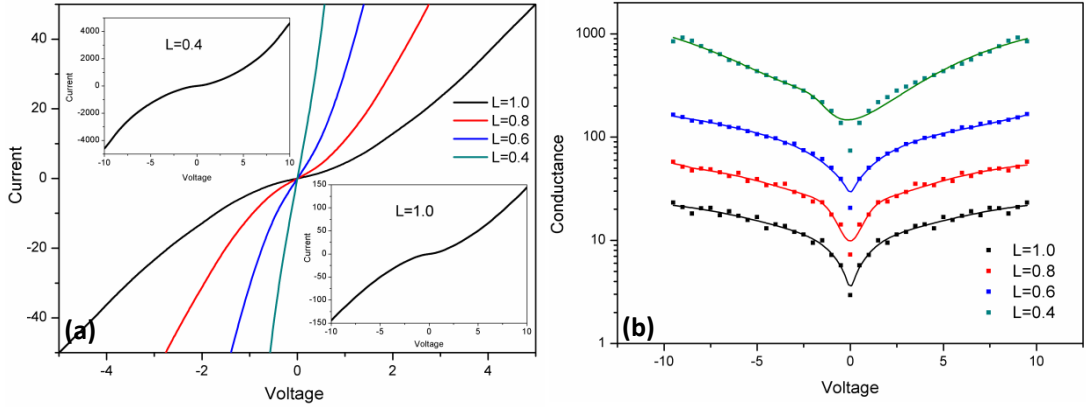

FIG. 11. (a) The *I-V* characteristics of the uniform rectangular array at different unit lengths for the temperature of 300 K. The upper inset shows the *I-V* curves of the array taken with a larger bias range at the unit length L=0.4. The lower inset shows the *I-V* curves of the array taken with a larger bias range at the unit length L=1.0. (b) The evolution of logarithmic plot of the differential conductance with the unit length at 300 K.

In this simulation we consider rectangular ordered arrays of 10×10 nanoparticles with different unit lengths. As shown in Fig. 11a, all of *I-V* curves are highly symmetric. The current reduces as the unit length increases. Examining the logarithmic plot of the respective differential conductances (see Fig. 11b) gives a deeper insight into the charge transport mechanism. The conductance is increasing when the unit length becomes shorter. When the tunnel distance between neighboring nanoparticles is decreasing, the length of barrier is becoming narrow and the distance between the nanoparticle and cathode is decreasing. So the transmission probability is increasing rapidly and the differential conductance is increasing when the unit length is becoming shorter. It is very important for the flexible nanoparticle array. It can adjust the value of conductance through the change of unit length in the experiment.

*4.1.3 The effect of local structural disorder on the characteristics of current-voltage in the 2D array*

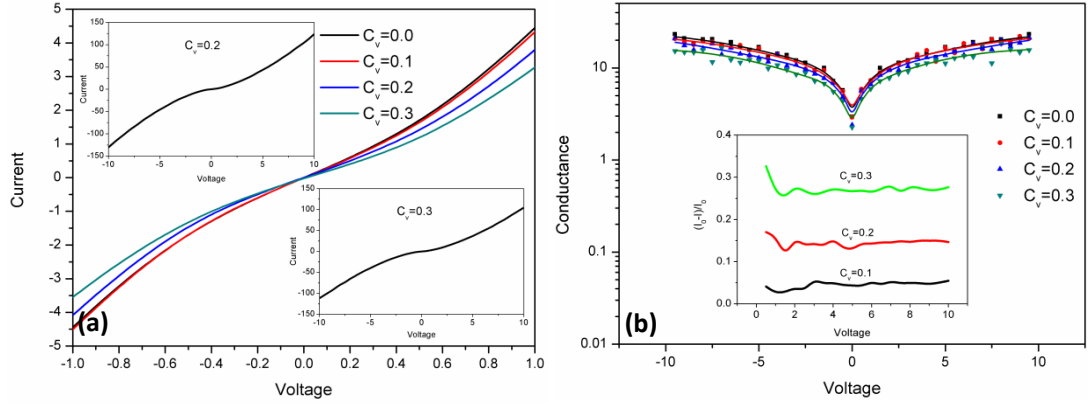

FIG. 12. (a) The *I-V* characteristics of rectangular arrays with different parameters $C_v$ when *s* is 1.0 at the temperature of 300 K. The upper inset shows the *I-V* curves of the same array taken with a larger bias range at the parameter $C_v$=0.2. The lower inset shows the *I-V* curves of the same array taken with a larger bias range at the parameter $C_v$=0.3. (b) The evolution of logarithmic plot of the differential conductance with the parameter $C_v$ for the same temperature. The lower inset shows $[(I_0-I)/I_0]$ versus voltage for different voltage directions. $I_0$ is the value of current when $C_v$ is zero.

The rectangular arrays of 10×10 nanoparticles with different parameters $C_v$ are used in this simulation at 300K. The unit length is $L_x = L_y = 1.0$. Fig. 12a shows the *IV* characteristics of four rectangular arrays with different variable parameters $C_v$ when *s* is 1.0 at 300 K. When $C_v$ is zero, it represents the ordered array without local disorder. However, the other arrays with nonzero $C_v$ represent disordered monolayers of monodisperse metal nanoparticles. The variable parameter $C_v$ shows the fluctuation of number density of nanoparticles, which means the appearance of new void in the film. From Fig. 12a, we can see that the *IV* characteristics of the nanoparticle arrays are nonlinear and highly symmetric whether the arrays with void or without void. However, the current reduces as the $C_v$ increases and the ratio of current change $(I_0-I)/I_0$ remain unchanged with the increasing of voltage, as shown in the lower inset of Fig. 12b. The logarithmic plot of the respective differential conductances is shown in Fig. 12b, which is decreasing with the improvement of parameter $C_v$.

When the variable parameter $C_v$ increases, the degree of local structural disorder of nanoparticle array increases. Numerous voids are created in the array, which leads to the variations on the tunnel distance. The tunnel distance therefore has a wide distribution. From the transmission probability, we can find the transmission probability inversely changed with the tunnel distance exponentially. When an electron transmits from one nanoparticle to the neighboring nanoparticle, due to the existence of void, the tunneling process cannot happen. So, the current and the differential conductance decrease with the improvement of $C_v$. As is shown in Fig. 12.

## 4.2 The characteristics of current-voltage in the 2D gradient array

*4.2.1 The effect of gradient on the characteristics of current-voltage in the 2D gradient array*

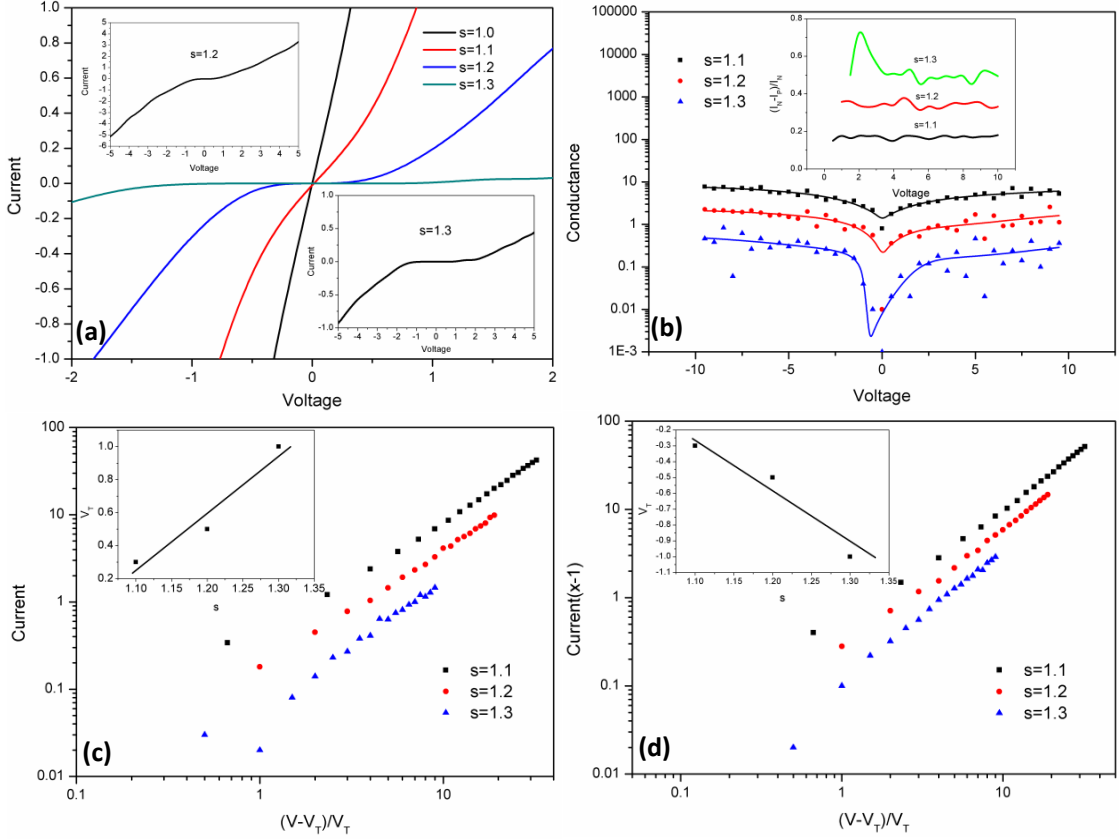

FIG. 13. (a) The *I-V* characteristics of rectangular gradient arrays with different factors *s* when $C_v$ is 0.0 at 300 K. The upper inset shows the *I-V* curves of the same array taken with a larger bias range at the parameter s=1.2. The lower inset shows the *I-V* curves of the same array taken with a larger bias range at the parameter s=1.3. (b) The evolution of logarithmic plot of the differential conductance with the parameter s for the same temperature. The upper inset shows [($I_N$ -$I_P$)/$I_N$] versus voltage for different factors *s*. $I_N$ is the absolute value of current when the applied voltage is negative. $I_P$ is the value of current when the applied voltage is positive. (c) The data is replotted as [(*V*-$V_T$)/$V_T$] versus current for positive voltage directions. The upper figure shows the threshold voltage versus parameter *s* in the positive voltage direction. (d) The figure is the power scaling [(*V*-$V_T$)/$V_T$] versus current for negative voltage direction. The inset shows the factor *s* dependence of the threshold voltage $V_T$ in the negative voltage direction.

The rectangular gradient arrays of 10×10 nanoparticles with different parameters *s* are used in this simulation at 300K. The unit length is $L_x = L_y = 1.0$. Fig. 13a shows the *IV* curves as a function of the parameter *s* for rectangular gradient arrays without local structural disorder. The curves are nonlinear and slightly asymmetric when the parameter *s* is not equal to 1. From the upper inset of Fig. 13b, we can see the degree of asymmetry of the current $(I_N - I_P)/I_N$ is increasing with the increasing the value of the parameter *s*, which indicates the asymmetry of the *IV* curve is dominated by the asymmetry of arrangement of nanoparticles. When the applied voltage is positive, electrons move from high number density region to low number density region. In the low number density region, the tunnel distance becomes larger and the electric potential decreases. The transmission probability will decrease exponentially. However, when negative voltage is applied, electrons move from low number density region to high number density region. Compared with the transmission probability of positive voltage, the transmission probability of negative voltage will be even larger. So, electron can flow easily from low density region to high density region.

The *IV* curves in Fig. 13 also exhibit a clear threshold voltage $V_T$. It is the characteristics of Coulomb blockade of electron transport. From the inset of Fig. 13c and the inset of Fig. 13d, the threshold voltage of positive applied voltage is the same as that in the negative applied voltage case, which indicates the largest tunnel distance plays a great role in the electron transport. And the threshold voltage fits a scaling behavior of the *IV* curves (see Fig. 13c and Fig. 13d).

*4.2.2 The effect of temperature on the characteristics of current-voltage in the 2D gradient array*

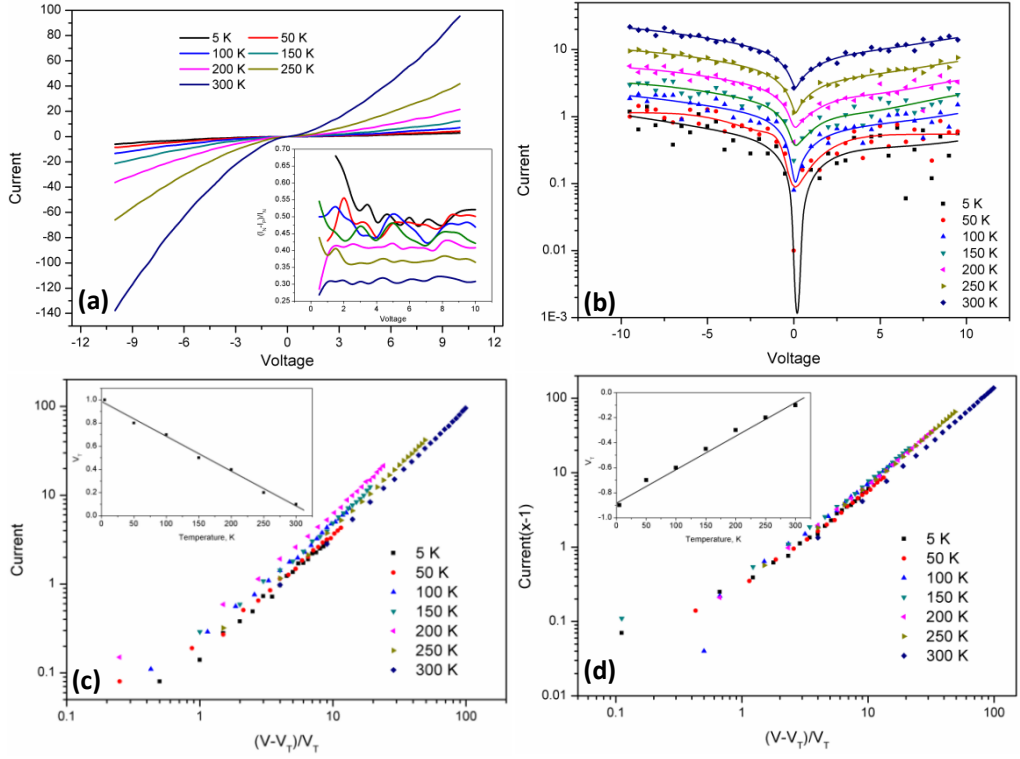

FIG. 14. (a) The *I-V* characteristics of rectangular gradient arrays with different temperature when $C_v$ is 0.0 and *s* is 1.3 at the unit length of 0.5. The lower inset shows $[(I_N - I_P)/I_N]$ versus voltage for different temperature. $I_N$ is the absolute value of current when the applied voltage is negative. $I_P$ is the value of current when the applied voltage is positive. (b) The evolution of logarithmic plot of the differential conductance with temperature. (c) The data is replotted as $[(V-V_T)/V_T]$ versus current for different temperatures. The upper figure shows the threshold voltage versus temperature in the positive voltage direction. (d) The figure is the power scaling $[(V-V_T)/V_T]$ versus current for negative voltage direction. The inset shows the temperature dependence of the threshold voltage $V_T$ in the negative voltage direction.

Rectangular gradient arrays of 10×10 nanoparticles with the unit length of 0.5 are used in this simulation at different temperatures. The *IV* characteristics calculated for a typical gradient array with $s=1.3$ are plotted in Fig. 14a for seven different temperatures. The *IV* curves were highly asymmetric, which can be seen from the degree of asymmetry of the current $(I_N - I_P)/I_N$. From the lower inset of Fig. 14a, we can see the degree of asymmetry of the current increases with the decreasing temperature. The logarithmic plot of the respective differential conductances is shown in Fig. 14b. The conductance is increasing with the improvement of temperature. For both positively and negatively applied voltage, the only adjustable parameter $V_T$ drops essentially linearly with increasing temperature (see the inset of Fig. 14c and the inset of Fig. 14d) and the *IV* characteristics are found to follow distinct scaling power law: $I \sim [(V - V_T)/V_T]^{\zeta}$ for the full temperature range we studied, 5-300 K (see Fig. 14c and Fig. 14d).

When temperature increases, the kinetic energy of electron $E_a$ increases and the energy gap $E_g$ ($E_g = V - E_a$, $V$ is the barrier height) decreases, so the transmission probability increases with temperature. Electron can have higher energy to overcome the barrier and then the current and the conductance increase as temperature rises. Therefore, the threshold voltage reduces with the increase of temperature. It belongs to the behavior of thermal electron emission.

### 4.3 The characteristics of current-voltage in the gradient array from 2D to 1D

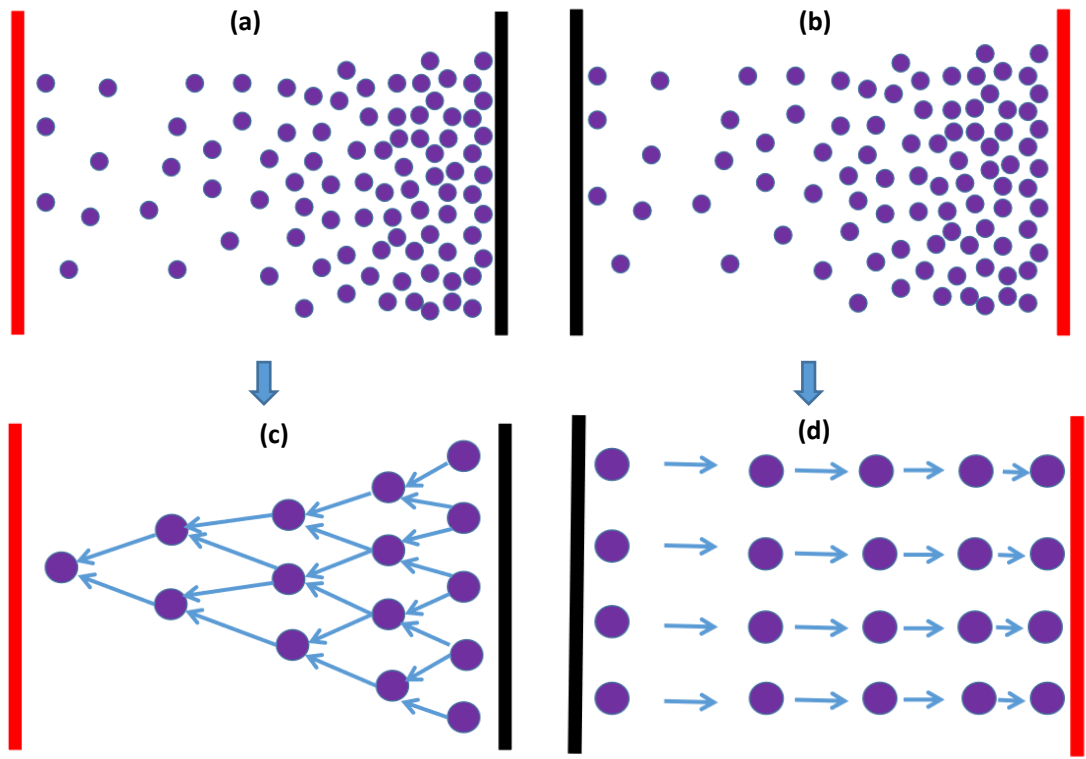

FIG. 15. (a) The schematic of gradient nanoparticle film, the electron is moving from high density region to low density region. (b) The schematic of gradient nanoparticle film, the electron is moving from low density region to high density region. (c) The process of electron moving from high density region to low density region. The model of triangular gradient array will be used to describe this process. (d) The process of electron moving from low density region to high density region. The model of rectangular gradient array will be used to describe this process.

We investigate the gradient nanoparticle array from 2D to 1D [40-45] by gradually decreasing number density of nanoparticle (see Fig. 15a). This will generate a certain gradient on the number density of nanoparticle. When electrons move from high number density region to low number density region, their flow paths change from 2D meandering to straight 1D channels, so the gradient array can be modeled on the triangular gradient array. However, when electron is moving from low number density region to high number density region, the gradient array can be modeled on the rectangular gradient array. Although the tunnel distance is very long in the low-density region, the distance between one nanoparticle and cathode is also significantly short. Electrons could pass through this huge barrier. In this section, rectangular gradient array of $N \times M$ nanoparticles ($N$=10, $M$=10) (see Fig. 15d) and triangular gradient array of $M$=10 (see Fig. 15c) are used to simulate the gradient array. Here we analyze the factors that influence the behavior of charge transport, such as temperature and gradient.

*4.3.1 The effect of gradient on the characteristics of current-voltage in the gradient array from 2D to 1D*

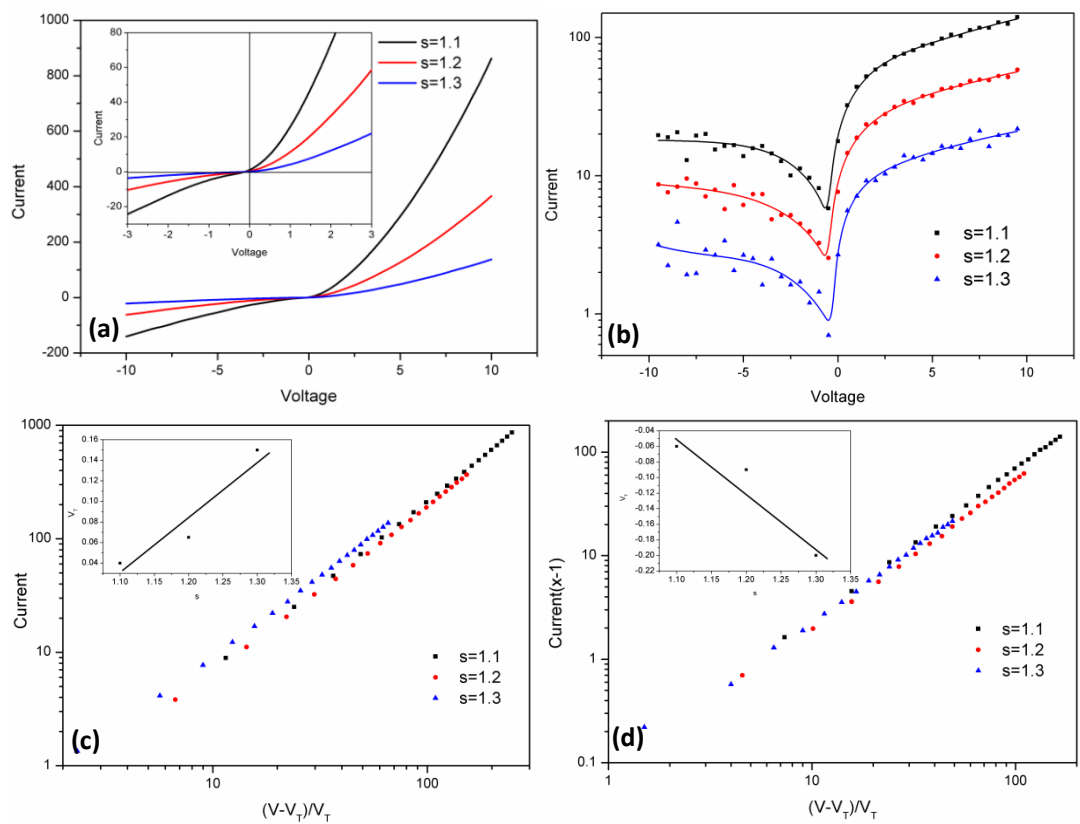

FIG. 16. (a) The *I-V* characteristics of 2D to 1D gradient arrays with different temperatures when $s$ is 1.3. The upper inset shows the *I-V* curves of the same array taken with a smaller bias range at the parameter s=1.3. (b) The evolution of the differential conductance with temperature. (c) The power scaling $[(V-V_T)/V_T]$ versus current in the positive voltage direction. The upper inset shows the temperature dependence of threshold voltage $V_T$. (d) The power scaling $[(V-V_T)/V_T]$ versus current in the negative voltage direction. The upper inset shows the temperature dependence of threshold voltage $V_T$.

In this simulation rectangular gradient array nanoparticles and triangular gradient array are used. The unit length is 0.5. Fig. 16a shows the *IV* curves as a function of parameter *s* for gradient arrays from 2D to 1D without local structural disorder. The curves are nonlinear and highly asymmetric, which indicates the asymmetry of the *IV* curve is dominated by the asymmetry of arrangement of nanoparticles (see Fig. 16a). The logarithmic plot of the respective differential conductances also shows the asymmetry of the conductance curve in Fig. 16b. The current and the conductance reduce as the parameter *s* increases. All arrays showed a clear voltage threshold for conduction, indicting strong Coulomb blockade behavior. For all arrays studied, $V_T$, the only adjustable parameter in the scaling procedure increases essentially linearly with the increasing the parameter *s*. The threshold voltage of positive applied voltage is very close to the threshold voltage of negative applied voltage (see the inset of Fig. 16c and the inset of Fig. 16d). For $V > V_T$, we plot current vs normalized voltage $v = (V/V_T - 1)$ on a log-log plot, as shown in Fig. 16c and Fig. 16d for two typical sets of data. We can see the linear asymptotic regime. When the parameter *s* increases, the tunneling distance becomes longer, so the transmission probability decreases, which leads to the reduction of current and the increase of the threshold voltage.

*4.3.2 The effect of temperature on the characteristics of current-voltage in the gradient array from 2D to 1D.*

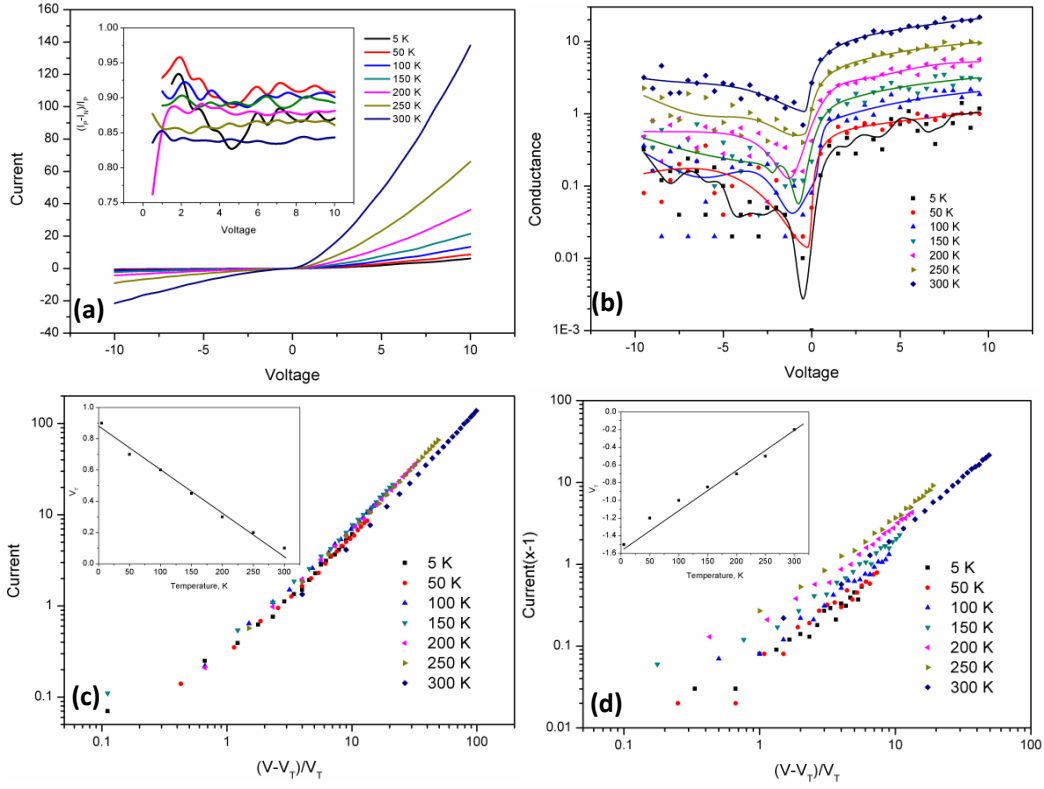

FIG. 17. (a) The *I-V* characteristics of 2D to 1D gradient arrays with different temperatures when *s* is 1.3. The upper inset shows the [$(I_P-I_N)/I_P$] versus voltage for different temperatures. $I_N$ is the absolute value of current when the applied voltage is negative. $I_P$ is the value of current when the applied voltage is positive. (b) The evolution of the differential conductance with temperature. (c) The power scaling [$(V-V_T)/V_T$] versus current in the positive voltage direction. The upper inset shows the temperature dependence of threshold voltage $V_T$. (d) The power scaling [$(V-V_T)/V_T$] versus current in the negative voltage direction. The upper inset shows the temperature dependence of threshold voltage $V_T$.

The *IV* characteristics calculated for a typical gradient array with $s=1.3$, which demonstrates a transition from 2D transport to 1D transport without local structural disorder, are plotted in Fig. 17a for seven different temperatures. The *IV* curves were highly asymmetric, which can be seen from the degree of asymmetry of the current $(I_P - I_N)/I_P$. From the upper inset of Fig. 17a, we can see the degree of asymmetry of the current increases with the decreasing temperature. For both positively and negatively applied voltage, the only adjustable parameter $V_T$ drops essentially linearly with increasing temperature (see the inset of Fig. 17c and the inset of Fig. 17d) and the *IV* characteristics are found to follow distinct scaling power law:

$I \sim [(V - V_T)/V_T]^\zeta$ for the full temperature range we studied, 5-300 K (see Fig. 17c and Fig. 17d).

When temperature increases, the kinetic energy of electron $E_a$ increases and the energy gap $E_g$ ($E_g = V - E_a$, $V$ is the barrier height) decreases, so the transmission probability increases with temperature. Electron can have higher energy to overcome the barrier and then the current increases as temperature rises. Therefore, the threshold voltage reduces with the increase of temperature. It belongs to the behavior of thermal electron emission and the asymmetrical $IV$ characteristics of array are very similar to the behavior of diode, which means such nanoparticle array could be used in the field of electronic device.

## 5. Conclusion

In this paper we have presented a study of the effects of the systematic variation of particle size distribution, a significant source of disorder within the system, on the temperature-dependent transport characteristics of the 2D metallic nanoparticle arrays with a new simulation method. Nanoparticle arrays with rectangular and triangular geometries are built to model the experimentally fabricated monolayers of nanoparticles. For a rectangular array with local structural disorder, the *I-V* curves are always non-linear and highly symmetric. However, the *I-V* curves are always non-linear and highly asymmetric in the rectangular gradient array and triangular gradient nanoparticle array. The results of simulation show that these nanoparticle arrays ranging from rectangular gradient array to triangular gradient array whether with local disorder or without local disorder show the clear voltage threshold due to Coulomb blockade and we find scaling in the current vs applied voltage for all the arrays, which is consistent with the prediction of MW [22]. The effect of temperature on the *IV* characteristics of two models is also investigated. The threshold voltage decreases linearly with temperature. It mainly affects the *IV* characteristics of metal nanoparticle arrays by random thermal motion.

In summary, lots of elements influencing the properties of electron transport in the array, such as number density of nanoparticle, temperature, size of nanoparticle, composition of nanoparticle and so on. Based on equation of the transmission probability, the tunnel distance is decided by number density of nanoparticle. So, the transmission probability is most sensitive with the tunnel distance. Electron can get more energy when the temperature is increasing, which can decrease the value of energy gap $E_g$. The transmission probability is more sensitive with temperature.

**Declaration of competing interest**

The authors declare that they have no known competing financial interests or personal relationships that could have appeared to influence the work reported in this paper.

**Acknowledgements**

This research was funded by the National Natural Science Foundation of China (Grant Nos. 11627806 and U1909214), the National Key R&D Program of China (Grant No. 2016YFA0201002).

**Author contributions**

Lianhua Zhang developed the ideas and theory for this work. Lianhua Zhang and the other authors prepared and edited the manuscript.